\documentclass[11pt,a4paper]{article}

\usepackage[utf8]{inputenc}
\usepackage[T1]{fontenc}
\usepackage[english]{babel}
\usepackage{lmodern}
\usepackage{geometry}
\usepackage{setspace}
\usepackage{amsmath,amssymb}
\usepackage{booktabs}
\usepackage{longtable}
\usepackage{array}

\usepackage[numbers,sort&compress]{natbib}

\usepackage{hyperref}
\hypersetup{colorlinks=true,linkcolor=blue,citecolor=blue,urlcolor=blue}

\title{\textbf{AI-Assisted Assessment of Experimental Physics Laboratory
    Reports: Potential, Limitations, and Support for Teaching Practice}}

\author{%
  Marcos Abreu$^{1}$,
  Cecilia Stari$^{2}$,
  Arturo C. Mart\'i$^{2,*}$\\[2pt]
  $^{1}$Consejo de Formaci\'on en Educaci\'on, Administraci\'on Nacional de
  Educaci\'on P\'ublica, Florida, Uruguay\\
  $^{2}$Instituto de F\'isica, Universidad de la Rep\'ublica, Montevideo,
  Uruguay\\}

\date{}

\begin{document}
\maketitle

\begin{abstract}
The incorporation of AI into the assessment of physics laboratory reports offers
opportunities to support the review of student work and the provision of
feedback. This article revisits an experience involving the use of ChatGPT
models to analyze laboratory reports and examines the conditions that influence
their usefulness in teaching practice. The analysis considers the evolution of
AI models, document-processing procedures, and different modes of interaction,
with particular attention to the effective availability of evidence contained in
student reports. The findings indicate that differences between AI-generated and
teacher assessments may be related not only to model capabilities but also to
difficulties in retrieving and interpreting equations, graphs, tables, units,
and other visual elements. The possibilities of systematic processing of sets of
reports and conversational review of cases requiring more focused analysis are
also examined. Based on these observations, a working strategy is proposed in
which AI serves as a complementary resource for organizing assessment,
supporting feedback, and identifying recurring difficulties in student work. The
analysis highlights the importance of document-processing conditions and
evidence availability, as well as the teacher's role in interpreting
AI-generated information and making evaluative and pedagogical decisions.
\end{abstract}

\textbf{Keywords:} Artificial intelligence in education; laboratory report
assessment; automated feedback; experimental physics.

\section{Introduction}

The incorporation of AI into assessment processes in higher education has raised
both opportunities related to automation and feedback and challenges associated
with the reliability of generated responses, teacher responsibility, and the
design of assessment practices \cite{xia2024scoping}. In particular, language
models can process student work and generate structured observations based on
previously established criteria. Their use to generate feedback in higher
education contexts has also begun to be explored specifically in physics
education \cite{mills2025prompting}. Within this context, an experience
conducted in the Experimental Physics course \cite{abreu2026exploring} explored
the use of AI models as support for the assessment and feedback of laboratory
reports. The automated analysis of written student work using natural language
processing techniques constitutes a relevant precedent for this type of
application in physics courses, particularly because of its potential to
systematically analyze large volumes of student work \cite{bralin2023analysis}.

Based on the observations that emerged from this experience, particular
attention was given to aspects related to the evolution of models, the different
ways in which documents can be processed, and the modes of interaction with
these systems. These issues made it possible to further examine the conditions
that may affect AI-generated results and the possibilities of integrating these
tools into different stages of teaching practice.

This article revisits this experience and analyzes some of the insights derived
from it, with particular attention to the possibilities and limitations
identified in the use of AI for laboratory report analysis. The purpose is to
provide elements for understanding the conditions under which these tools can
complement teachers' work and which aspects require professional teacher
intervention and judgment.

\section{Use of ChatGPT in the Assessment of Laboratory Reports}

There is previous research on the use of ChatGPT models for the assessment and
feedback of laboratory reports in the context of Experimental Physics courses
\cite{abreu2026exploring}. Likewise, tools based on generative models
specifically designed to provide formative feedback on physics students' work in
higher education have been explored \cite{mills2025prompting}.

In the study presented in \cite{abreu2026exploring}, reports corresponding to an
experimental activity on reaction time and statistics were analyzed. The reports
were initially assessed by instructors using a rubric and subsequently evaluated
using ChatGPT models according to equivalent criteria. The objective of this
experience was to analyze the possibilities offered by these models for
assessment and, particularly, for generating feedback on student work.

The analysis identified agreements and differences between the assessments made
by instructors and those generated by the model. Although the model demonstrated
the ability to analyze different components of the reports and formulate
structured assessments and comments, discrepancies were also identified in the
interpretation of some student submissions. These findings point to a central
issue for understanding the results obtained: model-generated responses depend
not only on the criteria used, but also on model capabilities and on how the
information contained in the reports is processed and made available to the
system.

\section{Evolution of Models and Report Processing}

The experience involving the assessment and feedback of laboratory reports
\cite{abreu2026exploring} included the use of different OpenAI models and
different processing approaches. This exploration made it possible to observe
how model capabilities and the ways in which information contained in documents
was accessed could influence the responses generated for the same task and for
different types of evidence.

The exploration included different generations of models, from GPT-3.5 and GPT-4
to GPT-4o and reasoning-oriented models. GPT-3.5 primarily supported tasks
involving textual review, reformulation of explanations, and identification of
straightforward conceptual errors. Subsequently, GPT-4 showed improvements in
contextual understanding and reasoning, supporting the analysis of consistency
among objectives, theoretical foundations, procedures, results, and conclusions
\cite{openai2023gpt4}. In 2024, GPT-4o introduced more developed multimodal
capabilities for jointly processing text and images, an aspect of particular
interest in reports containing graphs, tables, diagrams, and experimental images
\cite{openai2024gpt4o}. In parallel, reasoning-oriented models expanded the
possibilities for addressing multistep problems and mathematical and scientific
tasks \cite{openai2024reasoning}.

The evolution continued with GPT-5, GPT-5.1, GPT-5.2, and GPT-5.4, models that
were used at different stages of this work to explore their performance in the
assessment of laboratory reports. These versions incorporated progressive
improvements in reasoning, multimodal understanding, document analysis,
contextual understanding, and the ability to follow complex instructions. In
particular, GPT-5.4 was the most recent version used in the study
\cite{abreu2026exploring,openai2026gpt54}. The use of these different
generations made it possible to observe variations in the responses produced by
the models, both in the interpretation of student work and in the form and level
of detail of the assessments and feedback provided.

The evolution of models is particularly relevant in the case of laboratory
reports, since these productions integrate conceptual, mathematical,
experimental, and visual information. However, greater reasoning or multimodal
processing capabilities do not guarantee that all elements contained in a
document will be adequately available to the model. For this reason, in addition
to model capabilities, attention was given to how reports were processed and to
what information could actually be retrieved.

A central methodological consideration concerned the processing of reports in
PDF format. Sending a document directly does not guarantee that all of its
elements will be correctly recognized and interpreted. Possible difficulties
include the loss or modification of numbers, symbols, units, mathematical
expressions, graph labels, table structures, and relationships between different
elements on a page. These difficulties are particularly relevant in physics
laboratory reports, where a small modification to a numerical value, unit, sign,
or mathematical expression can affect the physical interpretation of a result.
Recent studies show that OCR systems and multimodal models continue to present
limitations when dealing with visually complex documents, and that high
character-recognition accuracy does not necessarily guarantee the preservation
of document structure and meaning
\cite{sun2026ocr,xu2026multilingual,khanal2026reforMe,khanchandani2026invoice,li2026mllms}.

In this regard, a fundamental methodological distinction should be made: the
fact that evidence is present in a report does not necessarily mean that such
evidence is effectively available to the model during processing. An equation
may be correctly included in the document but may not be adequately retrieved in
the processed content. Similarly, a graph may contain relevant information that
cannot be correctly interpreted if its axes, units, or legends are not legible.
This distinction is particularly important in the assessment of laboratory
reports, where evidence may be distributed across different formats.
Consequently, a difference between the model's assessment and that of an
instructor does not necessarily result from a difficulty in reasoning about the
physics content; it may also be related to the information that the model was
able to retrieve from the document. Document processing therefore constitutes a
relevant stage of the assessment process, since it can condition the information
on which the model subsequently operates.

Based on the processed information, assessments and feedback were generated
according to the criteria established in the rubric. Overall, the experience
showed that the usefulness of AI for laboratory report assessment depends on
multiple factors. It is not related solely to the model's reasoning
capabilities, but also to the quality of document processing and the ability to
adequately retrieve and interpret relevant evidence. These observations show
that document processing constitutes a relevant stage because it conditions the
information on which the model can subsequently operate. However, evidence
availability is not the only factor influencing the analysis. The way in which
the interaction with the model is established and the review of student work is
requested is also relevant.

\section{Different Ways of Interacting with Models}

The use of AI models to support assessment can be implemented through different
modes of interaction. In particular, API-based batch processing and
conversational interaction present different advantages and limitations, and
their usefulness therefore depends on the purpose of the task and the degree of
teacher intervention expected.

API-based batch processing offers as its main advantage the possibility of
assessing a large number of reports systematically, using the same instructions
and criteria for all submissions. This promotes procedural consistency and makes
it possible to rapidly generate an initial stage of feedback for a large group
of students. Automation also facilitates the organization of responses into
comparable structures and enables subsequent analysis of the results obtained.
In large-course contexts, this approach can reduce the time devoted to routine
tasks and allow instructors to focus their attention on cases requiring more
detailed review.

From the perspective of resource use, API-based processing also makes it
possible to control and plan the consumption associated with each request. Model
usage costs depend, among other factors, on the number of tokens processed.
These constitute units of text processing and may correspond to complete words,
parts of words, numbers, or symbols. Both the information sent to the model and
the generated response contribute to total usage. In the assessment of reports,
this usage depends on document length, the instructions provided, the length of
the rubric, and the amount of feedback requested. When large numbers of reports
are processed, these variables can become considerably important. Even a
relatively small difference in the number of tokens required to process each
report can accumulate significantly when the analysis is performed on dozens or
hundreds of submissions, increasing total token consumption and, consequently,
the resources required for processing. For this reason, the design of
instructions and the output format is also relevant to procedural efficiency.
Requesting excessively long responses, for example, may increase consumption
without necessarily producing a proportional improvement in their usefulness to
the instructor.

However, automated processing through an API also presents limitations. It
requires instructions, the rubric, and the output format to be defined in
advance, which can make the procedure less flexible when dealing with particular
situations that were not anticipated in the protocol.

Conversational interaction, in contrast, allows for more flexible and focused
interaction. The instructor can ask the model to review a specific criterion,
explain the basis of an observation, or examine a particular element of the
report. This possibility is especially useful when the initial assessment
produces a doubtful, superficial, or invalid response, as it allows the
interaction to be directed toward the evidence requiring additional
verification. In the experience analyzed \cite{abreu2026exploring}, this
approach made it possible to explore cases in which certain equations or figures
had not been reliably retrieved during batch processing.

From the perspective of resource consumption, conversational interaction behaves
differently. Each additional exchange may increase the amount of processed
tokens, particularly when information from the report must be provided again,
clarifications are added, or more detailed explanations are requested. This
increase does not represent a direct economic cost for the instructor; however,
it involves a greater processing volume and may affect available usage limits
and the time required to complete the task. For this reason, additional
interactions should preferably be used in a focused manner, particularly in
cases where the initial response requires further review or verification. Among
its disadvantages is the fact that it requires greater teacher involvement and
is therefore less efficient when the objective is to analyze large numbers of
reports simultaneously. In addition, the response may depend on the formulation
of the questions and the information provided during the interaction. The
formulation of instructions is therefore a relevant component of interaction
with these systems, particularly when the goal is to generate feedback on
student work \cite{wan2024exploring}. Consequently, conversational interaction
does not eliminate the need to verify AI-generated responses.

From a teaching perspective, both modes can be considered complementary. The API
is particularly suitable for an initial systematic review of a large set of
reports, whereas conversational interaction can subsequently be used to
investigate specific cases in greater depth. Thus, a hybrid workflow could
consist of initially processing all reports through the API, identifying
responses that present difficulties in justification or evidence access, and
subsequently conducting a focused review through conversational interaction.
This combination makes it possible to take advantage of the processing and
organizational capabilities of the API without giving up the flexibility of
dialogue.

However, neither mode should be interpreted as a mechanism for replacing teacher
judgment. The API and conversational interaction can provide information,
suggestions, and preliminary feedback, but the instructor is responsible for
verifying the evidence, contextualizing the observations, and determining
whether they are appropriate for assessment and feedback.

\section{Guidance for Teachers}

The use of AI systems as tools to support the assessment of laboratory reports
raises several considerations regarding their incorporation into educational
contexts. These recommendations are not intended to establish an autonomous
grading procedure, but rather to guide the incorporation of AI as a
complementary resource that can help organize the review, generate an initial
analysis of reports, and identify aspects requiring expert judgment.

The usefulness of AI is not necessarily uniform across the different criteria of
a rubric. Its incorporation is therefore more appropriate when it is established
in advance which tasks the system can support, which evidence it should
retrieve, and in which situations teacher intervention is required.

\subsection{Use of AI}

A fundamental aspect of incorporating AI is to understand it as a support tool
that helps organize and focus the review, while professional teacher judgment
guides the interpretation of evidence and the final assessment decision.

The experience analyzed shows that, even when the model uses the same criteria
as the rubric employed by the teaching staff, AI-generated scores are not
interchangeable with teacher grades. Consequently, the system output should be
considered an input for guiding the review rather than a definitive grade.

AI can produce a first structured reading of the report, identify possible
strengths and difficulties, and organize information according to the rubric
criteria. The instructor can subsequently use this information to confirm,
modify, or reject the observations made by the model. This approach makes it
possible to take advantage of AI's ability to process large amounts of
information without transferring responsibility for assessment decisions to the
system.

The analysis shows that AI-generated feedback presents different levels of
reliability depending on the type of evidence involved. Criteria related to more
descriptive aspects may provide useful information for an initial review,
whereas those requiring interpretation of equations, graphs, units,
uncertainties, or relationships among different parts of the report require more
careful teacher verification.

For this reason, a useful strategy is to employ AI to identify where teacher
attention should be concentrated rather than to automatically determine what is
correct or incorrect. Feedback can function as a guide indicating:

\begin{itemize}
  \item which rubric criteria appear to be adequately addressed;
  \item which aspects present insufficient or unclear evidence;
  \item which statements require direct verification in the report;
  \item which assessments may be affected by extraction or interpretation
    problems;
  \item which students or groups present recurring difficulties that may require
    pedagogical intervention.
\end{itemize}

In this way, AI can contribute to prioritizing teachers' work and reducing the
time devoted to repetitive tasks without replacing the pedagogical
interpretation of student work.

\begin{table}[htbp]
  \centering
  \caption{Guidance for using AI feedback by rubric item. Authors' own
    elaboration based on the data from the analyzed experience.}
  \label{tab:guia_ia}
  \begin{tabular}{p{0.30\textwidth}p{0.62\textwidth}}
    \toprule
    \textbf{Rubric item} & \textbf{Recommendation for the instructor}\\
    \midrule
    Objectives             & May be accepted after general review; this is the
                             safest item to delegate to AI.\\
    Theoretical framework  & Manually review the feedback, especially formulas
                             and mathematical notation.\\
    Experimental setup     & Verify comments referring to images or experimental
                             apparatus diagrams before accepting them.\\
    Data analysis          & Prioritize teacher review of calculations, units,
                             graphs, and uncertainties.\\
    Conclusions            & Compare with the graphical evidence cited in the
                             report before validating the score.\\
    Overall assessment     & Use only as an orienting input.\\
    \bottomrule
  \end{tabular}
\end{table}

\subsection{Rubric and Instruction Design}

The use of AI is more consistent when assessment criteria are formulated in
observable and verifiable terms. It is recommended to:

\begin{itemize}
  \item Write rubric criteria in the most observable and verifiable way
    possible, avoiding formulations that can only be assessed through an overall
    reading of the report.
  \item Explicitly require the model, in the instructions, to support each score
    with a concrete reference to the report text (citable evidence), thereby
    avoiding generic validation responses such as ``the criterion is present.''
  \item Include a structured checklist for each item, replicating, as far as
    possible, the same checks that would be applied by the teaching team.
  \item Ask the model to explicitly state when it cannot verify an aspect
    because of a lack of legible evidence, rather than completing the response
    with a plausible but unsupported interpretation.
\end{itemize}

These conditions can improve the clarity and evidential basis of the feedback,
making it easier for the instructor to quickly verify the evidence supporting
the observations made by the AI.

\subsection{Mitigating Limitations}

One of the main limitations identified in this experience concerns the retrieval
of information contained in equations, figures, tables, and graphs. Extraction
errors may generate plausible but insufficiently supported feedback or even
invalid assessments. For this reason, instructors should pay particular
attention to AI observations that depend on:

\begin{itemize}
  \item mathematical expressions;
  \item numerical values and units;
  \item graphs and their labels;
  \item data tables;
  \item experimental diagrams;
  \item images incorporated into the report.
\end{itemize}

To promote adequate interpretation of these elements, the following practices
are recommended:

\begin{itemize}
  \item Ask students to submit reports as high-resolution PDF files, avoiding
    low-quality scanned or photographed images.
  \item Recommend that equations be written using the word processor's equation
    editor rather than inserted as images, since fractions and symbols embedded
    as images may be difficult to read automatically.
  \item Require reports to include graphs with clearly legible axes, units, and
    legends within the figure itself. In addition to being necessary for proper
    interpretation, this requirement facilitates information retrieval during
    automated processing.
  \item When feedback is classified as doubtful or invalid, prioritize a second
    review in conversational mode focused on the specific criterion and evidence
    rather than discarding the tool altogether.
\end{itemize}

\subsection{A Working Strategy}

Based on the experience analyzed, a combined workflow is proposed for large
courses. AI can perform an initial systematic review of reports using the
rubric, generating observations organized by criterion. Subsequently, the
instructor can prioritize cases in which the system indicates difficulties in
verification, difficulties in grounding the assessment in verifiable evidence,
or possible inconsistencies.

The process can be organized into three stages:

\begin{enumerate}
  \item \textbf{First AI-assisted review:} batch processing of reports through
    an API and generation of structured feedback.
  \item \textbf{Identification of cases requiring attention:} detection of
    superficial or invalid assessments, or assessments supported by evidence
    that presents difficulties in access or interpretation.
  \item \textbf{Focused teacher review through conversational interaction:}
    verification of the evidence in the original report and, when necessary,
    interaction with the model in conversational mode to request a new analysis,
    point to a specific equation, table, graph, or fragment of the report, and
    direct the model's attention toward that evidence. Based on this review, the
    instructor can prepare, supplement, or modify the final feedback.
\end{enumerate}

This strategy makes it possible to concentrate teachers' time on aspects
requiring greater disciplinary and pedagogical expertise, using AI for an
initial homogeneous review and conversational interaction to investigate cases
requiring specific analysis.

\subsection{Using AI to Identify Patterns and Guide Teaching}

In addition to supporting the review of individual reports, AI can be used to
identify trends and recurring difficulties across sets of submissions. When an
instructor receives a large number of reports, individual review can reveal the
errors present in each submission, but it is more difficult to quickly identify
which difficulties recur across several students.

In this context, joint processing of the reports can provide a complementary
view of group performance. For example, analysis of a set of reports may reveal
that a particular difficulty occurs repeatedly across different submissions.
These difficulties may be related to the treatment of uncertainties and
significant figures, the interpretation and construction of graphs, the
selection and use of physical models, the handling of units, the description of
experimental procedures, the communication of results, or the formulation of
conclusions based on the evidence obtained. Identifying these patterns can help
instructors distinguish isolated errors from difficulties requiring broader
pedagogical intervention.

The information obtained can also be organized according to the different
criteria of a rubric. In this way, rather than focusing exclusively on
individual student performance, instructors can obtain an overall representation
of the aspects in which the group presents the greatest difficulties. For
example, if a substantial proportion of reports presents problems with graph
interpretation, this information can guide the incorporation of specific
activities on data representation, reading, and interpretation. Similarly, if a
recurring difficulty is observed in the formulation of conclusions, it may be
appropriate to explicitly address the relationship between experimental results,
evidence, and conclusions.

This type of processing can also contribute to collective feedback. Based on the
patterns identified, instructors can select some of the most frequent errors or
difficulties to address in a common session, avoiding the need to repeat the
same explanations individually across numerous reports. AI can therefore assist
in preparing summaries of the main difficulties identified and organizing
examples that can be discussed with the group.

However, the results of this analysis should be interpreted as supporting
information for teaching decisions rather than as an automatic diagnosis of
learning. In this sense, the main contribution of AI does not necessarily lie in
producing a grade, but in providing systematized information that helps
instructors interpret students' collective performance and make informed
pedagogical decisions. Processing multiple reports can thus contribute to
identifying teaching priorities, planning reinforcement activities, selecting
content requiring greater attention, and designing collective feedback sessions.
AI is thereby incorporated as a tool that supports the teacher's perspective and
can facilitate the identification of patterns that might be difficult to
recognize through the exclusively individual review of a large volume of work.

\section{Conclusions}

The use of AI to analyze laboratory reports cannot be reduced to a comparison
between an AI-generated grade and a teacher's assessment. The possibilities for
its use depend on a set of conditions that include model capabilities, document
processing, the effective availability of evidence, and the mode of interaction.

From this perspective, the incorporation of artificial intelligence can be
understood as a process that allows for different forms of support for teaching
practice. Systematic processing can facilitate the initial organization of a
large set of student submissions, while conversational interaction makes it
possible to investigate particular cases and review specific evidence in greater
depth. In turn, information obtained from multiple reports can contribute to
recognizing recurring difficulties and guiding subsequent teaching decisions.

In this context, the main contribution of these tools does not necessarily lie
in producing an automated assessment, but in expanding the possibilities for
analyzing student work and facilitating certain aspects of teaching practice.
The interpretation of evidence, evaluation of results, and pedagogical decisions
require consideration of the context of each submission and remain connected to
teachers' professional knowledge. Consequently, regardless of the evolution of
generative artificial intelligence, it can be understood as a resource that can
be flexibly integrated into different stages of assessment and teaching,
provided that attention is paid to the conditions under which information is
generated and interpreted.

\bibliographystyle{unsrtnat}
\bibliography{referencias}

\end{document}